\documentclass[twocolumn,floatfix,preprintnumbers,citeautoscript]{revtex4} %,floats
\usepackage{graphicx}
\usepackage{amsmath,amssymb,graphics}
\usepackage{bbm} 
\usepackage{accents} 
\usepackage{mathrsfs}	% e.g. nice Lagrange-L with \mathscr{L}
\usepackage[T1]{fontenc}

\newcommand{\E}[1]{\ensuremath{\mathrm{E}_{#1}}} % e.g\E{8}
\newcommand{\G}[1]{\ensuremath{\mathrm{G}_{#1}}}
\newcommand{\SO}[1]{\ensuremath{\mathrm{SO}(#1)}}
\newcommand{\SU}[1]{\ensuremath{\mathrm{SU}(#1)}}
\newcommand{\U}[1]{\ensuremath{\mathrm{U}(#1)}}
\newcommand{\Z}[1]{\ensuremath{\mathbbm{Z}_{#1}}} % Z_N ->\Z{N}
\newcommand{\BmL}[0]{\ensuremath{B\!-\!L}}
\DeclareMathOperator{\Tr}{Tr}

\begin{document} \setlength{\unitlength}{1in}

\preprint{ CERN-PH-TH/2007-050; DESY 07-030; TUM-HEP-660/07}
 
\title{\Large{Seesaw Neutrinos from the Heterotic String}}

\author{{\bf\normalsize Wilfried~Buchm\"uller$^1$\!,
Koichi~Hamaguchi$^{2}$\!,
Oleg~Lebedev$^{3}$\!,}\\
{\bf\normalsize
Sa\'ul~Ramos-S\'anchez$^4$\!,
Michael~Ratz$^5$}\\[0.5cm]
{\it\normalsize
${}^1$ 
Deutsches Elektronen-Synchrotron DESY, 22603 Hamburg, Germany}\\[0.15cm]
{\it\normalsize
${}^2$ Department of Physics, University of Tokyo, Tokyo 113-0033, Japan
}\\[0.15cm]
{\it\normalsize
${}^3$ CERN, Theory Division, CH-1211 Geneva 23, Switzerland}\\[0.15cm]
{\it\normalsize
${}^4$ Physikalisches Institut der Universit\"at Bonn,}\\[-0.05cm]
{\it\normalsize Nussallee 12, 53115 Bonn,
Germany}\\[0.15cm]
{\it\normalsize
${}^5$ Physik Department T30, Technische Universit\"at M\"unchen,}\\[-0.05cm]
{\it\normalsize James-Franck-Strasse, 85748 Garching, Germany}
}

\begin{abstract} 
We study the possibility of realizing the neutrino seesaw mechanism in the
$\E8\times \E8$ heterotic string. In particular, we consider its $\Z6$ orbifold
compactifications leading to the supersymmetric standard model gauge group and
matter content. We find that these models possess all the necessary ingredients
for the seesaw mechanism, including the required Dirac Yukawa couplings and 
large Majorana mass terms. We argue that this situation is quite common in
heterotic orbifolds. In contrast to the conventional seesaw of grand unified
theories (GUTs), no large GUT representations are needed to generate the
Majorana mass terms. The total number of right--handed neutrinos can be very
large, up to $\mathcal{O}(100)$.
\end{abstract}

\pacs{\dots}

\maketitle

The seesaw mechanism \cite{Minkowski:1977sc,Yanagida:1979ab,GellMann:1979cd}  is perhaps the most 
attractive way
to explain the smallness of the neutrino masses.
Its essential ingredients are  heavy Majorana neutrinos and their Yukawa
couplings to the left--handed neutrinos. The supersymmetric seesaw mechanism is
described by the superpotential
\begin{equation}\label{eq:seesawW}
 W~=~Y_\nu^{ij} \,\phi^u\,\ell_i\, N_j + \frac{1}{2} M_{jk} N_j\, N_k \;,
\end{equation}
where $\phi^u$ and $\ell_i$ ($i=1,2,3$) are the Higgs and lepton doublets, and $N_j$
($1\le j\le n$) are some heavy standard model singlets.  At low energies, this
leads to 3 light neutrinos with masses of order $(Y_\nu\, \langle \phi^u \rangle )^2
/ M$,  where $Y_\nu$ and $M$ represent typical values of $Y_\nu^{ij}$ and
$M_{jk}$, respectively. For $Y_\nu \sim 1$ and $M\sim 10^{16}\,\mathrm{GeV}$,
one has $m_\nu \sim 10^{-3}\mathrm{eV}$.  The scales of the atmospheric and solar
neutrino oscillations, $\sqrt{\Delta m^2_\mathrm{atm}}\simeq0.04\,\mathrm{eV}$
and $\sqrt{\Delta m^2_\mathrm{sol}}\simeq0.008\,\mathrm{eV}$
\cite{Maltoni:2004ei}, are suspiciously close to this scale. This hints at GUT
structures behind the seesaw. 

In conventional GUTs, $N_j$ are members of GUT matter multiplets, e.g.\
a $\boldsymbol{16}$--plet of \SO{10}, and $M_{jk}$ are related to a  vacuum
expectation value (VEV) of a large GUT representation, e.g.\ a 
$\overline{\boldsymbol{126}}$--plet of \SO{10}.  In this case the Majorana mass
terms originate from the coupling 
$\boldsymbol{16}\,\boldsymbol{16}\,\overline{\boldsymbol{126}}$ 
(cf.~\cite{Mohapatra:2006gs}). 
 
Difficulties arise when one attempts to embed this attractive mechanism in
string theory \cite{Witten:1985bz}, which is the best candidate for the 
ultraviolet completion of the standard model. One of the problems is that 
string GUTs usually do not allow for large GUT representations such as 
$\overline{\boldsymbol{126}}$. Other obstacles
are posed by string selection rules for the superpotential couplings which
forbid some of the necessary couplings in certain models
\cite{Giedt:2005vx}. 
For some string models one can obtain Dirac
neutrino masses \cite{Ibanez:2001nd,Antoniadis:2002qm}. Recently, large 
Majorana masses have been obtained from instanton effects 
\cite{Blumenhagen:2006xt,Ibanez:2006da,Cvetic:2007ku} 
as well as from higher order operators \cite{Kim:2004pe}
and small Majorana masses from
large volume compactifications \cite{Conlon:2006wt}. However, it turned
out to be very difficult to obtain the standard seesaw mechanism in
consistent string models which lead to the gauge group and the matter
content of the minimal supersymmetric standard model (MSSM). 

The purpose of this letter is to show that these problems can be circumvented in
the $\E8\times \E8$ heterotic string \cite{Gross:1984dd,Gross:1985fr}. We base
our study on orbifold compactifications of the heterotic string
\cite{Dixon:1985jw,Dixon:1986jc,Ibanez:1986tp,
Ibanez:1987sn,Casas:1987us}.
Recent work on an orbifold GUT interpretation of heterotic models 
\cite{Kobayashi:2004ud,Forste:2004ie,Kobayashi:2004ya} has triggered renewed
interest in these constructions. It has lead to the idea  of  stringy local
grand unification
\cite{Buchmuller:2005jr,Buchmuller:2004hv,Buchmuller:2005sh,Buchmuller:2006ik},
which allows for an intuitive explanation of the multiplet structure of the
standard model as well as gauge coupling unification. 

The key point of local grand unification is that fields localized in the compact space   at points with
a GUT symmetry appear as complete representations of that GUT in the low--energy
spectrum. This applies, in particular, to $\boldsymbol{16}$--plets of local
\SO{10} GUTs. On the other hand, the low--energy gauge group is an intersection
of  various local GUTs in $\E8\times\E8$  and can be just the standard model gauge group. 
It has been shown that, in the context of the \Z6-II orbifold, the
concept of local grand unification facilitates  construction of
phenomenologically promising models which admit vacua with the exact matter
content of the MSSM \cite{Buchmuller:2005jr,Buchmuller:2006ik,Lebedev:2006kn}.
These models inherit certain features of  the underlying GUTs 
such as the GUT structure of matter multiplets and the standard GUT hypercharge.
There are  also  significant   differences from the usual 4D GUTs. Most
importantly, the bulk and higher twisted sector states appear in split
multiplets, thus allowing for natural doublet--triplet splitting. Furthermore,
mixing of the localized generations with vector--like bulk  states breaks  the unwanted
GUT relations for the fermion masses \cite{Asaka:2003iy,Buchmuller:2006ik}.

\begin{figure}[!h!]
\centerline{\includegraphics[width=5cm]{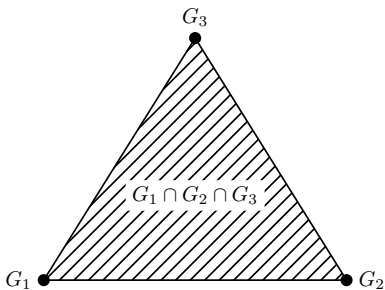}}
\caption{Couplings involving fields localized at different fixed points in 10D 
preserve the symmetry corresponding to the intersection  of the local gauge groups $G_i$.
\label{fig:gilg}}
\end{figure}

The Yukawa couplings do not necessarily preserve the symmetry of local GUTs.
The symmetry of the non--local  coupling in 10D is an intersection
of the local gauge groups at the vertices (Fig.~\ref{fig:gilg}).
This implies, for example, that the Majorana mass terms for the 
neutrino components of the  $\boldsymbol{16}$--plets can
originate from the coupling 
\begin{equation}
 \nu_{\boldsymbol{16}}\times\nu_{\boldsymbol{16}}
 \times(\text{SM singlets})\;,
\end{equation}
where the singlets belong neither to $\accentset{(-\!\!-\!\!-)}{\boldsymbol{16}}$ nor
to $\accentset{(-\!\!-\!\!-\!\!-\!\!-)}{\boldsymbol{126}}$ 
of  \SO{10}. 
Furthermore, any SM singlet can play the role of the right--handed neutrino as long
as it has a Yukawa coupling to the lepton doublets and a large Majorana mass. These are abundant in orbifold
models and typical models contain  $\mathcal{O}(100)$ such singlets.

Heterotic  
orbifold models  are defined by the orbifold twist $v$, torus lattice
and the gauge backgrounds which include the gauge shift $V$ and the Wilson lines
$W$.
With an appropriate choice of these quantities, one can obtain
models with 
(i) the standard model gauge group  $G_\mathrm{SM}$  times extra group factors, and
(ii) 3 matter generations, 2 Higgs doublets plus extra vector--like
states \cite{Buchmuller:2005jr,Lebedev:2006kn}. 
Such models contain
an `anomalous' U(1) which induces the Fayet-Iliopoulos (FI) $D$-term,
\begin{equation}
D_\mathrm{anom}~=~\frac{g\,M_\mathrm{Pl}^2}{192 \pi^2}~ \Tr \mathsf{t}_\mathrm{anom} 
+ \sum_i q_i\, \vert \phi_i \vert^2 \;,
\end{equation}
where $\mathsf{t}_\mathrm{anom}$ is the anomalous U(1) generator, $q_i$ are the
anomalous charges of fields $\phi_i$ and $g$ is the gauge coupling. This
triggers spontaneous gauge symmetry breaking while preserving supersymmetry
\cite{Dine:1987xk}. Some of the fields charged under the anomalous U(1) (and, in
addition, under other gauge groups) develop non--zero VEVs  thereby  reducing 
gauge symmetry.  The scale of these VEVs is set by the FI term which is somewhat
below the string scale. This  eventually determines the seesaw scale.  In
general, any SM singlets can get large VEVs as long as it is consistent with
supersymmetry, and one can obtain the standard model gauge  symmetry times that
of the hidden sector, 
\begin{equation}
 G~\longrightarrow ~\SU{3}_c\times\SU{2}_\mathrm{L}\times \U1_Y 
 \times G_\mathrm{hidden} \;.
\end{equation} 
The singlet VEVs are not necessarily associated with  flat directions in the
field space and generally  correspond to isolated solutions to supersymmetry equations \cite{Buchmuller:2006ik}.
The hidden matter gauge group  $G_\mathrm{hidden}$ can be responsible
for spontaneous supersymmetry breaking. In fact, within the class of
models with the MSSM spectrum,
gaugino condensation
in the hidden sector favours   TeV--scale soft masses
for the observable fields \cite{Lebedev:2006tr}.

The non--zero singlet VEVs lead to the mass terms for the vector--like states,
\begin{equation} 
 W~=~ x_i\, \bar x_j\, \langle s_a\, s_b\, \dots \rangle \;,
\end{equation}
where $x_i , \bar x_j$ are the vector--like exotics and $\langle s_k \rangle$ are
the SM singlet VEVs in string units. Such a coupling must be consistent with the string 
selection rules \cite{Hamidi:1986vh,Dixon:1986qv}. It has been shown
that many \Z6-II  models satisfy this requirement and all of 
the vector--like exotics can be decoupled  \cite{Lebedev:2006kn}. 
 This results in the MSSM spectrum at low energies.

Similarly, the singlet VEVs induce ``Majorana'' mass terms for the SM singlets
as well as the neutrino Yukawa couplings of Eq.~\eqref{eq:seesawW},
\begin{equation}
M_{ij} \sim \langle s_a\, s_b\, \dots \rangle ~~,~~ Y_\nu^{ij} \sim  \langle s_\alpha \, s_\beta\, \dots \rangle~,
\end{equation}
as long as it is consistent with string selection rules.

Identification of right--handed neutrinos is intimately related to the
issue of  baryon/lepton number violation.  In generic vacua, any SM singlet
can play the role of the right--handed neutrino. However, such vacua also
suffer from excessive R--parity violating interactions. 
The simplest way to suppress these interactions is to identify
a $\BmL$ gauge symmetry and enforce either its approximate conservation 
or conservation of its  discrete  (``matter parity'')  subgroup.
In local GUTs, the $\BmL$ generator  resembles the standard GUT $\BmL$,
but also requires  extra U(1) components beyond SO(10). It is non--anomalous
and produces the standard $\BmL$ charges for the SM matter.
If $\BmL$ is broken by VEVs of fields carrying even charges under $\BmL$,
the matter parity $(-1)^{3(\BmL)}$ is conserved. This forbids dangerous
 R--parity violating interactions and requires the right--handed neutrino
to carry the charge $q_{\BmL}=\pm 1$.
Another possibility is that  $\U1_{\BmL}$ is broken at an intermediate 
scale $M_{\BmL}$ such that all  R--parity violating couplings are
suppressed by $M_{\BmL}/M_\mathrm{Pl}$. In this case,  Majorana mass
terms for the right--handed neutrinos are allowed only upon
$\BmL$ breaking, which lowers  the seesaw scale to intermediate
energies. In what follows, we consider 
these possibilities in specific heterotic orbifold models.

{\it Model 1} \cite{Lebedev:2006kn}.  It is a $\Z6$-II orbifold based on a Lie torus lattice
 $\G2\times\SU3\times\SO4$  with a twist vector
 $v_6=\left(\frac{1}{6}, \frac{1}{3}, -\frac{1}{2} \right)$ (see \cite{Buchmuller:2006ik}
for details). 
The gauge shift in an  orthonormal $\E8 \times \E8$ basis is given by
\begin{eqnarray}
   V & = &
   \left(\tfrac{1}{3},-\tfrac{1}{2},-\tfrac{1}{2},0^5\right)\,
   \left(\tfrac{1}{2},-\tfrac{1}{6},(-\tfrac{1}{2})^5,\tfrac{1}{2}\right) \;, \nonumber
\end{eqnarray}
while the Wilson lines are chosen as 
\begin{eqnarray}
W_{2} & =
&\left(\tfrac{1}{4},(-\tfrac{1}{4})^4,(\tfrac{1}{4})^3\right)\,
\left(1,-1,-\tfrac{5}{2},-\tfrac{3}{2},-\tfrac{1}{2},-\tfrac{5}{2},-\tfrac{3}{2},\tfrac{3}{2}\right) \;, \nonumber \\
W_{3} & = &\left(-\tfrac{1}{2},-\tfrac{1}{2},(\tfrac{1}{6})^6\right)\,
\left(\tfrac{10}{3},0,-6,-\tfrac{7}{3},-\tfrac{4}{3},-5,-3,3\right)\;. \nonumber
\end{eqnarray}
Here the superscripts denote repeated entries.
The gauge group after compactification is $G_\mathrm{SM}\times\SO8\times\SU2\times\U1^7$,
while the matter content is that of the MSSM plus vector--like exotics.
 The $\BmL$ generator is identified with 
%\begin{equation}
\[
 \mathsf{t}_{\BmL}~=~
 \left(
 1 , 1 , 0 , 0 , 0 , -\tfrac{2}{3} , -\tfrac{2}{3} , -\tfrac{2}{3} \right)\,
 \left(
 \tfrac{1}{2} , \tfrac{1}{2} , 0 , \tfrac{1}{2} , \tfrac{1}{2} , 0 , 0 , 0
 \right)\;.
% 
% \mathsf{t}_{\BmL}~=~\left(1,\,1,\,0,\,0,\,0,\,-\tfrac{2}{3},\,-\tfrac{2}{3},\,-\tfrac{2}{3}\right)\left(2x - \tfrac{1}{2},\,\tfrac{1}{2},\,0,\,x,\,x,\,0,\,0,\,0\right) \;,
%\end{equation}
\]
%where $x$ is arbitrary.
In general supersymmetric configurations, many SM singlets get non--zero VEVs. Choosing 
a subset of such singlets with 0 or $\pm 2$  $\BmL$ charges, the  unbroken gauge symmetry is 
\begin{equation}
G_\mathrm{SM} \times G_\mathrm{hidden} \;,
\end{equation}
where $ G_\mathrm{hidden}= \SO8$, while all of the exotic states get large masses and decouple.
This vacuum preserves the matter parity $(-1)^{3(\BmL)}$.

We find that there are 39 right--handed neutrinos   defined by $q_{\BmL}= \pm
1$, two of which are members of the localized  $\boldsymbol{16}$--plets. They 
have Yukawa couplings to the lepton doublets and large Majorana mass terms. We
have calculated the $3\times 39$ Yukawa matrix  $Y_\nu$ and    $39\times 39$
Majorana mass matrix  $M$ of Eq.~\eqref{eq:seesawW} up to 6th order in the
singlet VEVs.  
That is, for each matrix element, we have determined at which order 
in the superpotential a non--zero coupling is allowed by string selection rules.
Each entry depends on the quantum numbers and the localization of 
the  Majorana neutrinos, and involves products of different singlets 
and moduli--dependent Yukawa couplings.
We then assume that the main hierarchy in these entries comes 
from products of singlet VEVs so that these matrices can be treated 
as textures.
The effective mass matrix for the left--handed neutrinos,
\begin{equation}
 M_\mathrm{eff}~=~ - v_u^2\,Y_\nu\, M^{-1}\, Y_\nu^T 
\end{equation}
with $v_u$ being the up--type Higgs VEV, 
can be represented  by the texture  
\begin{equation}
M_\mathrm{eff}~\sim~ - \frac{v_u^2}{M_*}  \left( 
\begin{matrix}\label{texture}
1 & s & s \\
s & s^2 & s^2 \\
s & s^2 & s^2 
\end{matrix}
\right) \;.
\end{equation}
Here $s < 1$ represents a generic singlet VEV in string units 
and $M_*$ is the effective seesaw scale. 
$Y_\nu$ contains entries with powers of $s$ between 1 and 5, while 
  the dependence of the eigenvalues of $M$  ranges from
$s$ to $s^8$
(with no massless eigenstates at generic points in moduli space). 
This results in a strong $s$-dependence of the effective seesaw
scale $M_*$. This scale is 
further suppressed by the large multiplicity of heavy
singlets $N$, $M_* \propto N^{-x}$ with $0<x<2$. The value of $x$ depends
on the texture. For example, when all the singlets contribute equally,
$x=2$, whereas $x=0$ if only a fixed number of neutrinos have
non--negligible couplings.  
For the present model, we find
\begin{equation}
 M_* ~\sim~ 0.1\, s^5\, M_{\rm str} ~\sim~ 10^{14}\,\mathrm{GeV}\;,
\end{equation}
for the string scale $M_{\rm str} = 2\cdot 10^{17}\,\mathrm{GeV}$ and $s\sim 0.3$.
The obtained texture \eqref{texture} is of course model dependent.

The corresponding charged lepton Yukawa matrix is of the form 
\begin{equation}
Y_e~\sim~ \left( 
\begin{matrix}
 1 & 1 & c \\
 c & c & c^2 \\
 0 & 0 & 0 
\end{matrix}
\right) \;,
\end{equation}
where ``0'' denotes absence of the coupling up to 6th order in the singlet VEVs
$c$. Such zeros are expected to be filled in at higher orders. Here we are again
using a single expansion parameter although in  practice there are many
variables.

These crude estimates show that reasonable fermion masses can in principle
be obtained. Inserting order one coefficients in the textures, one finds that the
eigenvalues scale as
\begin{equation}
m_{\nu_i}~\sim ~(1, s^2, s^2)\,\frac{v_u^2}{M_*}\;,\quad
 m_{e_i}~\sim~(1,c,0)\,v_d \;,
\end{equation}
where $v_d$ is the down--type Higgs VEV.
For $s\sim 0.3$ and $c\sim 0.1$ the textures reproduce roughly the observed
lepton mass hierarchy. 
The above texture favours the normal neutrino mass
hierarchy and can accommodate small and large mixing angles.
Further details of the model are available at \cite{seesaw-website:2007ws}.

{\it Model 2} \cite{Buchmuller:2005jr}. It is based on the same orbifold
with a different choice of  the gauge backgrounds.
The gauge shift and the Wilson lines are
 \begin{eqnarray}
 V & = &
 \left(\tfrac{1}{2},\tfrac{1}{2},\tfrac{1}{3},0,0,0,0,0\right) \, 
 \left(\tfrac{1}{3},0,0,0,0,0,0,0\right) 
 \;,\nonumber\\
 W_2 & = & 
 \left(\tfrac{1}{2},0,\tfrac{1}{2},\tfrac{1}{2},\tfrac{1}{2},0,0,0\right) 
 \,\left(-\tfrac{3}{4},\tfrac{1}{4},\tfrac{1}{4},-\tfrac{1}{4},\tfrac{1}{4},\tfrac{1}{4},\tfrac{1}{4},-\tfrac{1}{4}\right) 
 \;,\nonumber\\
 W_3 & = &
 \left(\tfrac{1}{3},0,0,\tfrac{1}{3},\tfrac{1}{3},\tfrac{1}{3},\tfrac{1}{3},\tfrac{1}{3}\right) \, 
 \left(1,\tfrac{1}{3},\tfrac{1}{3},\tfrac{1}{3},0,0,0,0\right)
 \;. \nonumber
 \end{eqnarray}
The $\BmL$ generator is given by  
%\begin{equation}
\[
 \mathsf{t}_{\BmL}~=~
 \left(
 0,1,1,0,0,-\tfrac{2}{3},-\tfrac{2}{3},-\tfrac{2}{3}
 \right)
 \,
 \left(
 \tfrac{1}{2},\tfrac{1}{2},\tfrac{1}{2},-\tfrac{1}{2},0,0,0,0
 \right)
 \;.
\] 
%\end{equation}

This model also produces the exact MSSM gauge group and spectrum  (up to the hidden sector)
in generic SUSY vacua where many SM singlets get large  VEVs.
Interestingly, there are 21 SM singlets with $q_{\BmL}=+1$ and 
18 singlets with $q_{\BmL}=-1$. 
We found that their $21\times 18$ mass matrix $M_{ij}$ has the full rank once singlets 
with $q_{\BmL}=0$ develop VEVs. Thus, there are exactly three (massless) right--handed neutrinos with $q_{\BmL}=+1$
in the limit of $\U1_{\BmL}$ conservation.

Unlike model 1, model 2 does not have a natural R-- or matter parity once the 
$\U1_{\BmL}$ symmetry is broken. This is  because there are no singlets with 
even  nonzero  $q_{\BmL}$. The simplest way to suppress R--parity violation  is to break  
$\U1_{\BmL}$ at an intermediate scale. 
This occurs when the SM singlets charged under $\BmL$ develop
hierarchically small VEVs, $M_{\BmL} \ll M_\mathrm{str}$. 
(Although the details of this mechanism are not clear, this scale can, 
in principle, be generated dynamically and may be related to 
non-perturbative dynamics such as formation of condensates.)
The three right--handed neutrinos with $q_{\BmL}=1$, which remain massless under the conserved 
$\BmL$, now obtain masses of order $M_{\BmL}^2/M_\mathrm{str}$ via VEVs of singlets with $q_{\BmL}=-1$. 
They contribute to the neutrino
masses as
\begin{equation}\label{con1}
\Delta m_\nu \sim v_u^2 ~\frac{y^2}{{M_{\BmL}}^2/ M_\mathrm{str}}\;,
\end{equation}
where $y $ is the corresponding $\BmL$
conserving Yukawa coupling.

Furthermore, states with $q_{\BmL}\not=1$ can also play the role of 
right-handed 
neutrinos.  The corresponding  Yukawa couplings violate $\BmL$
and are suppressed by  $M_{\BmL}/M_\mathrm{str}$.
Their contribution is
\begin{equation}\label{con2}
\Delta m_\nu \sim v_u^2 \,\frac{(M_{\BmL}/M_\mathrm{str})^2}{\widetilde{M} }\;,  
\end{equation}
where $\widetilde{M}$ is the corresponding Majorana mass. 

Comparing Eqs.~\eqref{con1} and \eqref{con2}, we see  that
the first contribution induces  too large neutrino masses, unless
$y$ is suppressed, while the second contribution is very small, 
unless $\widetilde{M}$ is suppressed. Such suppression can come about when 
the relevant 
couplings are only allowed at high orders by string selection rules or if
the relevant singlet VEVs are very small. 
We have calculated Yukawa couplings $y$ for a class of vacua in model 2.
Although the necessary ingredients for the seesaw are present, 
we have not been able to identify a natural suppression mechanism for $y$
 when one links suppression of R--parity
violation  to an approximate $\U1_{\BmL}$.
The solution is likely to require going beyond our crude
approximation that many different singlets
attain similar VEVs.

Our  main conclusion  is that heterotic orbifolds allow for many realistic 
vacua with the massless spectrum of the MSSM \cite{Lebedev:2006kn}, which
also have the required ingredients for the seesaw. 
In such vacua, many SM singlets
get large VEVs consistent with supersymmetry (not necessarily along flat
directions). This generates Majorana masses for abundant SM singlets, 
which play the role of the right--handed neutrinos. The effective seesaw scale 
is model--dependent, but  typically  somewhat   below the 
string scale.
In models with local GUTs, this scale is intimately related to $\BmL$ breaking.
Although the structure of matter multiplets is similar to that of 4D GUTs, no
large GUT representations are required to generate  Majorana  mass terms. As a
result, potentially realistic neutrino mass patterns can be obtained.

\textbf{Acknowledgements.} 
This research was supported by the DFG cluster of excellence Origin and
Structure of the Universe, the European Union 6th framework program
MRTN-CT-2004-503069 "Quest for unification", MRTN-CT-2004-005104
"ForcesUniverse", MRTN-CT-2006-035863 "UniverseNet" and SFB-Transregios 27
"Neutrinos and Beyond" and 33 "The Dark Universe" by Deutsche
Forschungsgemeinschaft (DFG).

\end{document}